# A Set of Essentials for Online Learning : CSE-SET

J. Dulangi Kanchana (Corresponding author) [1] , Gayashan Amarasinghe[1] , Vishaka Nanayakkara[1] and Amal Shehan Perera[1]

[1] Department of Computer Science and Engineering,
Faculty of Engineering,University of Moratuwa, Sri Lanka..
dulangik@cse.mrt.ac.lk, gayashan@cse.mrt.ac.lk, vishaka@cse.mrt.ac.lk,
shehan@cse.mrt.ac.lk

**Abstract.** Distance learning is not a novel concept. Education or learning conducted online is a form of distance education. Online learning presents a convenient alternative to traditional learning. Numerous researchers have investigated the usage of online education in educational institutions and across nations. A set of essentials for effective online learning are elaborated in this study to ensure stakeholders would not get demotivated in the online learning process. Also, the study lists a set of factors that motivate students and other stakeholders to engage in online learning with enthusiasm and work towards online learning.

## 1    Introduction

With the advancements in technology, students now have easy access to learning management systems through their smart mobile devices, making online learning more convenient [1]. In contrast to traditional classroom learning, which is teacher-centric, online education takes a student-centric approach [2]. Therefore, successful online learning requires a significant amount of self-directed effort [3],[4]. If a student is motivated, they can engage in lectures and complete exercises. However, while there is some facilitator interaction, students cannot be compelled to attend lectures, and thus, it is crucial for students to be motivated to participate in class discussions.

According to a study on online education in Greece, instructors used Internet-sourced instructional materials to teach their pupils. They posted additional announcements and instructional information but not on online platforms or relevant Learning Management Systems. The study concluded that motivation among stakeholders is crucial for effective online learning to occur [5].

Figure 1 depicts Maslow's hierarchy of needs and Alderfer's ERG theory. The elements of these theories can have an impact on the behavior of learners and play a crucial role in enhancing their learning experience. To participate in online lectures or even physical education, a student must have their fundamental physiological needs met, which includes housing, clothes, food, clean water, clean air, and access to a functional computer/mobile device, power source, and internet connectivity. Safety needs for online education include the institution's promise to the student that their information will not be misused, and this also applies to students' and parents' anxieties about the potential misuse or bullying that could occur with internet access [6] [7]. Social needs include a sense of belonging to a particular academic year or cohort of students, the sense of community among classmates or colleagues, and involvement with academic staff with shared conversation themes.

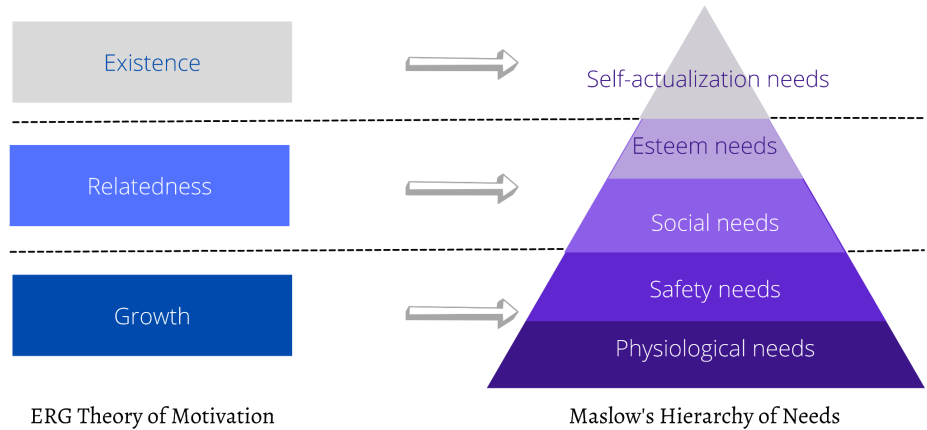

**Fig. 1.** ERG Theory of Motivation and Maslow's Hierarchy of Needs (McLeod, 2007; Alderfer, 1969) respectively.

The students' esteem needs are fulfilled through the evaluation of their lecture attendance. This evaluation system is designed to encourage more students to attend class. Students who attend 100% or more than 80% of the lectures receive five additional points on their final subject grade. This positive reinforcement technique is used to promote desirable behavior. In contrast, students with less than 50% attendance have five points deducted from their final subject grade. This approach utilizes negative reinforcement to prevent undesirable conduct.

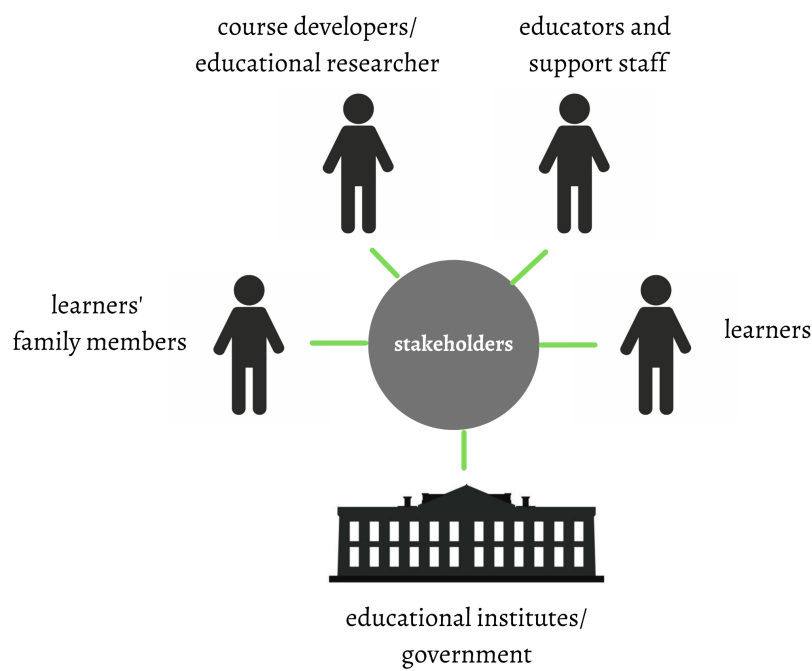

**Fig. 2.** Stakeholders involved in the learning process (Romero & Ventura, 2010) respectively.

Fig. 2 portrays the various stakeholders involved in education, including students, educators, parents, course developers, educational researchers, educational institutes, and system administrators [8]. In cases where the student still resides with their parents, the financial support provided by the parents should be considered, along with their positive or neutral attitude towards online education, as opposed to a negative or insecure mindset. It is also important to ensure that parents are financially stable enough to provide computers/mobile devices, electricity, and internet access.

Computer-mediated communication can enhance students' information acquisition in physical classrooms, thereby contributing to their cognitive growth [9]. Collaborations and partnerships within universities can also promote teaching quality and research standards and even improve the infrastructure of universities [10].

Effective online learning requires interaction between teachers and students, appropriate teaching styles, and the provision of knowledge-rich information. Group tasks are often used to enhance peer collaboration. Course design, the knowledge obtained, and the ability to apply knowledge to future activities are essential factors from a student's perspective [11]. Providing assistance to learners in completing course-related tasks in an online learning environment can increase student satisfaction with the course [12].

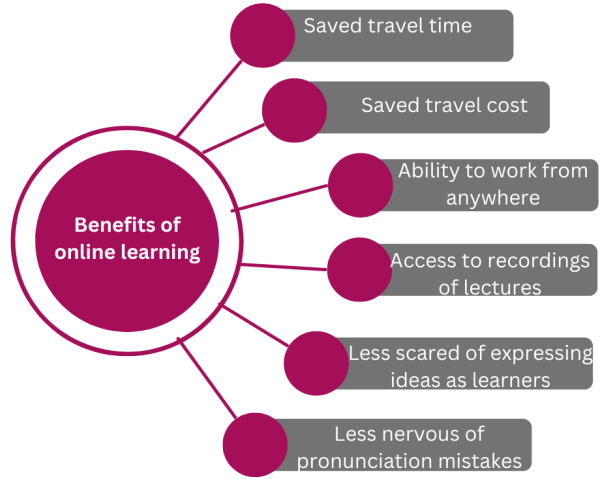

Fig. 3. Benefits of online learning.

Online learning offers several benefits, including the elimination of the need to travel to and from a physical classroom, which saves students valuable time that can be used for other purposes. Moreover, online learning can reduce transportation costs for students who no longer need to commute to a physical location to attend class. The flexibility of online learning enables students to work or study from the comfort of their own homes, allowing them to better balance their personal and professional commitments. Recorded lectures, which are often available on online learning platforms, provide students with the opportunity to review material as many times as they need to fully understand it. Online learning can also offer a less intimidating environment for learners, reducing self-consciousness about pronunciation errors and creating a more comfortable and relaxed learning environment. Fig. 3 illustrates the benefits of online learning.

Online learning provides students with the opportunity to obtain foreign degrees, regardless of their geographical location. Even students in rural areas can benefit from online learning without the need to attend a physical educational institution in a city. With the increasing popularity of online learning, academic institutions are also adopting this mode of education.

Despite its growing popularity, concerns remain about the effectiveness of online learning compared to traditional learning methods. This raises questions about how online learning can be made more effective. In this article, our aim is to improve the effectiveness of online learning by addressing the following research questions:

a) What are the essential factors required for students and educators to participate in online learning?
b) What are the factors that are necessary to motivate students for a positive learning experience in online learning?

The insights gained from research in this area can help teaching staff adopt better teaching styles and strategies for delivering online courses through learning management systems. Additionally, the knowledge gained through such studies can aid in the design of more effective online courses, resulting in a better learning experience for learners. Such research can also foster innovation and development in online learning platforms, enabling them to better meet the needs of learners and teaching staff.

This article aims to make online learning a more effective experience and addresses several research questions. The approach section of this paper outlines the review methodology, while Section 3 presents the key findings on the essential factors necessary for the proper and effective conductance of online learning. Finally, Section 4 concludes the paper.

## 2    Approach

 For this review, we considered education and learning in online modes for the higher education target sector. Papers written in other languages than English were excluded. Only peer-reviewed articles with a primary focus on online learning or those that discussed some aspects of online learning were included in this study. Non-peer reviewed articles were excluded from the study. The journal articles, workshop papers, and conference papers written in the last 20 years were retrieved by searching databases: ACM Digital Library, IEEE Xplore, ScienceDirect, Scopus and SpringerLink using *Self-Regulated Learning, Self-Directed Learning, Online education, Distance education, Online learning, Motivation, Student satisfaction, e-learning, Technology enhanced learning* as keywords.

## 3    Key findings

Based on the existing literature have been identified as essential components needed for online learning: Connected, Self-regulated, Environment, System usability and maintenance, Engagement and Technical fluency.

The key findings have been arranged in the following order as mentioned in the paragraph: connectedness, self-regulation, engagement, system usability and maintenance, environment, and technical fluency.

### 3.1    Connected

Some of the components of online learning comprise live streaming of lecture sessions where both learners and academic staff participate. To enable this, they must have:

    A.  Possession of/access to a functional computer or mobile device
    B.  Accessibility to an uninterrupted electricity supply
    C.  Stable internet connection

Access to high-speed internet enables a broader audience to receive education [18]. However, online learning has been proven to be challenging for students residing in areas with poor internet connectivity and unreliable power supply [19]. In order to effectively engage in online learning, a consistent power supply and stable internet connection are essential prerequisites for any region, district, or country [20][21].

To overcome the challenges posed by poor internet connectivity, teachers or facilitators can upload lecture sessions for students to download and view later if they are unable to attend the live session. In addition, lectures can be divided into shorter-duration videos to enable students to download and view them with ease [13]. Providing lecture videos in both video and audio formats also helps students to access the learning materials using devices that are most convenient for them [23].

However, a lack of access to functional computer equipment is another barrier to effective online learning [22]. To overcome this challenge, educational institutions or the government can provide loan facilities for students who need to acquire computer equipment [21]. It is also important to note that some learning management systems (LMS) may not support mobile devices, thereby limiting access to the resources required for online learning [23].

For students in remote areas without access to the internet, distance education through television or radio transmission may be an option [21]. Academic staff may also require additional resources to facilitate online education, such as portable internet dongles with a predetermined data allowance. Alternatively, they may be provided with additional allowances to acquire the necessary resources for online education [24].

Academic institutions can acquire or lend computer equipment to their faculty, or the employees may be instructed to acquire the materials with the institution covering the cost of acquisition. In families where both the parent and child require access to a computer for online education, the institution may need to provide additional resources to accommodate scheduling conflicts between the child's online learning session and the parent's online lecture session.

In summary, ensuring access to reliable power and a stable internet connection, providing access to functional computer equipment, and offering distance education options are critical factors that can help improve the effectiveness of online learning.

## 3.2 Self-regulated

In the online learning process, learners are mostly responsible for their own learning, and therefore, they need to possess self-regulated attributes and behaviours that reinforce self-regulatory behaviours, such as:

A. Self-discipline towards studies
B. Persistent nature of the learner
C. Self- directed effort of student

    D.   Self- evaluation of student
    E.   Self-satisfaction of the learner

Self-regulation is an essential characteristic of humans for achieving objectives [25]. Students with greater self-regulation utilise ways to optimise the learning process [26][27] in the education sector.

Due to their self-regulatory behaviour, students' motivation is boosted [26]. Students can be motivated to succeed or accomplish their objectives. Students can attain desired learning outcomes, skills, or competencies [28]. Acquiring new information will be an accomplishment for learners and a source of their joy. Also, the student will develop good self-confidence as a result of their work [29]. By being self-motivated, individuals can take control of their learning and become more efficient and effective learners. Such learners are more likely to set goals, plan their activities, monitor their progress, and adjust their behaviors based on feedback.

Self-discipline is the ability to control one's own behavior and overcome distractions, temptations, and impulses that may interfere with achieving goals. Self-motivation can lead to the development of self-discipline behaviors, which are important for success in online learning and other areas of life.

These favourable characteristics will be developed in students who adapt to online learning systems. However, students who have a more substantial commitment to an office job or home duties and receive less family support are less likely to adjust to education than students who receive family assistance [30]. Students who are unable to successfully adapt to an online learning environment are unlikely to demonstrate continuous efforts toward reaching educational goals [37][30].

Significant learning happens when learners comprehend newly taught information, connect it to their past knowledge, and apply the combined knowledge to practical day-to-day life events and new circumstances [31][32]. Here, explicit information is transformed into tacit understanding. The term for this process is externalisation [33]. While lecturers and professors can utilise narratives to relate their personal or professional experiences alongside theory taught to help students recall and comprehend abstract topics taught [34].

When students express their frustration in attempting to solve challenging problems with other students, their motivation is increased since they realise that this is a frequent occurrence for everyone. Peer interaction is, therefore, a source of motivation in collaborative learning environments (Moller, 1998). Student frustration and dissatisfaction are caused by internet connectivity issues in online learning [35]. Even in the context of physical education, peer collaboration contributes to a positive university experience and can decrease the likelihood of students dropping out [36].

Technological is a tool for online learning [37][38], but the student's ownership of the technology component does not necessarily indicate motivation. However, the absence of technology to engage in online learning might result in dissatisfaction.

Therefore, technology is a hygiene component of online learning. Consequently, according to Maslow's Need Theory, technology may be viewed as a physiological demand.

In online learning, learners have more autonomy and responsibility, and they need to be able to regulate their own learning behaviors, such as setting goals, managing their time, monitoring their progress, and adjusting their learning strategies. Students may postpone their learning if they receive recorded lectures and lengthy deadlines for completing continual assessments. Since recordings are readily accessible at any time, learners may put off attending learning events and engaging in learning activities [39].

Even in the post-pandemic situation, when learning shifted back to traditional classroom learning, some students still expect lecturers to record the session and allow hybrid learning mode, a combination of traditional classroom and online learning [39]. Closer to the examinations, students may be overburdened with work and lose the time and capacity to finish their assignments [41]. With physical education, students might sense the competitiveness of their peers. However, it is still being determined whether this is the case in online education. Self-regulated learning also helps learners develop metacognitive skills, such as self-reflection and self-evaluation, which are essential for lifelong learning. Self-regulatory behaviours, self-discipline, and self-motivation are crucial to the practical completion of courses and course modules for students enrolled in online education or learning.

Engagement in CSE-SET framework or peer-to-peer or teacher-to-peer contact is a motivating factor and, according to Maslow's Need Theory, social needs. Interaction in whatever form is a source of motivation. Hence, according to Herzberg's theory, interactivity is a motivating factor.

### 3.3    Engagement

Learning is a two-way process, and interactions with peers and academic staff are crucial in facilitating learning outcomes. Let us examine the significance of interaction and engagement in online education.

    A.    Students with proper discipline who do not disrupt the class
    B.    The students' interaction level with the teacher or teachers
    C.    Peer interactivity
    D.    The concentration level of an average learner
    E.    The teacher's style of instruction
    F.    Presence of a discussion forum

Determining whether students are actively engaged in the learning process or simply attending live online lectures while participating in other activities, such as playing games, can raise reasonable doubts among instructors. It is also possible to question whether students or third parties are completing continual assessments. In online learning environments, measuring students' learning process is a challenging task due to the lack of information regarding student learning behavior and achieved learning outcomes. Furthermore, cheating and plagiarism are prevalent among students in online learning contexts [61].

Adult students who participate in live online lecture sessions should act responsibly to avoid disturbing other participants or the lecturer. Technical support provided to students to promote online learning can enhance their course satisfaction. The instructor's years of expertise in the classroom, random questioning based on learners' attention levels, sharing prior experiences with students, and using videos to explain abstract theoretical concepts can help keep students engaged. The teaching style of the instructor also plays a significant role. If the instructor is understandable, learners are more likely to participate, and vice versa [42].

As students interact more with their instructors and classmates, their course satisfaction increases [62]. In traditional classroom settings, students have the chance to interact with their classmates and teachers face-to-face, which helps to develop a sense of community and collaboration. However, in online learning, this sense of community can be difficult to establish due to the distance and lack of physical presence. Discussion forums, video conferencing, and group assignments and social media platforms can facilitate student interaction in online learning and education.

Interacting with other students enrolled in the online course can help clarify subject-related doubts. Peer interaction is a critical aspect of both in-person and online education. Discussion forums enable the instructor and other students to respond to student inquiries [63]. Peer interaction, which involves two-way contact between students to exchange knowledge, ideas, and opinions, can lead to the application and assessment of learning and reflections on learning.

Collaborative development of beliefs and meaning through conversations between individuals in a group facilitates learning [64]. Learners who participate in online learning activities often encounter difficulties, such as lack of clear guidance from educators and ineffective teaching strategies [60].

A lecture can be either teacher-centric or student-centric. Student-centered learning allows students to share their experiences and actively engage in learning. The duration of an online session should be determined based on the typical time that human concentration can be maintained.

## 3.4    System usability and maintenance

The satisfaction of learners is positively influenced by the user-friendliness, reliability, high availability, and portability of the LMS, as well as its ability to achieve the intended capabilities of the LMS [65]. High availability of the LMS refers to its ability to be accessible at all times by both students and teachers [66].

Assigning staff, such as programmers and system administrators, to maintain the LMS is crucial for online educational support [66]. Appropriate measures should be adopted to ensure the confidentiality of students' data. The educational institution should provide the necessary infrastructure, as described. As shown in Fig. 4, according to Herzberg's two-factor theory, poor hygiene factors reduce learner satisfaction, while motivation factors increase learner satisfaction.

However, in specific fields such as medicine, learning materials may be limited [61]. When online examinations are conducted, they are open-book exams where students have access to study materials and Google search options. Due to the shorter time allowed to complete an online exam compared to a physical exam, online exams may be more challenging [21].

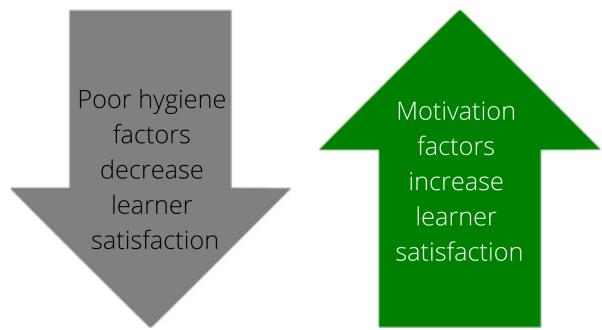

**Fig. 4.** Stakeholders involved in the learning process (Romero & Ventura, 2010) respectively.

### 3.5    Environment

Students require a peaceful environment to study, while teaching staff also require a calm environment with no disturbances to effectively deliver instruction. Here, we will examine the importance of the environment attribute in online education :

    A.    Enough space for online learning activities and studying
    B.    No sound disturbance
    C.    Has sufficient time for studies
    D.    Support from family towards education

If students are the primary providers for their families, they may have limited time for studying [56]. Students' motivation to engage in online learning can also be influenced by their families' attitudes towards online education. If students' parents hold unfavorable views about online learning, students may be less likely to participate in it. Conversely, if students' families encourage them to engage in online learning, it can boost their motivation [57]. Additionally, parental goal-setting behavior can enhance students' motivation and achievement in education [57].

In physical education, where most students reside in university residences near campus, peer motivation to attend physical lectures plays a crucial role. However, in online learning, recorded lecture sessions are available to students, and they can access them at their convenience. This flexibility and convenience may prompt some students to choose online learning over traditional classroom learning [58].

Learners require an environment with adequate lighting, a comfortable temperature, and minimal noise disruption [59]. Those who are accustomed to studying in libraries may find their homes comparatively less spacious [43]. As illustrated in Fig. 4, hygiene factors are essential for physical and online learning to occur without dissatisfaction. However, these factors do not always indicate that a student is motivated to learn. Spending long hours in front of a computer screen can cause eye fatigue for both instructors and students. To prevent this, teachers should schedule breaks every hour to allow educators and students to rest their eyes [13].

Conflicts within the learner's family or relationships can hinder their learning and studying processes [60]. In physical education, students who live near campus and away from home have the opportunity to avoid family conflicts. A stress-free environment is crucial for students to concentrate on their academics. It is also important for instructors to have a distraction-free environment when conducting live or recorded sessions to avoid disturbing their learners [60].

The surrounding atmosphere is also crucial for academic personnel. Instructors who join a live or recorded session from a noisy background can disturb learners. Therefore, it is essential for instructors to have a distraction-free environment to deliver lectures as intended. Family disputes should be kept to a minimum to allow instructors to concentrate on providing effective instruction.

### 3.6    Technical fluency

The delivery of online lectures requires a distinct set of skills compared to traditional classroom teaching methods. However, instructors may encounter challenges in comprehending that successful video production requires time and effort to ensure the content remains focused, concise, and engaging for learners. It is not advisable to rely solely on a written script to convey the message as it may appear unnatural and not conducive to an effective learning experience. Additionally, instructors may require several prior sessions to become comfortable with being on camera, particularly if they experience camera shyness [42].

**Table 1.** Popular Learning Management Systems.

| Learning Management System | References |
|---|---|
| Google Classroom | [13][14] |
| BigBlueButton | [15] |
| GoToMeeting | [15] |
| Blackboard | [15] |
| Open edX | [16] |
| Moodle | [14][15][16] |

Table 1 presents a list of popular learning management systems used by educational institutions. Before commencing online-based teaching, instructors, lecturers, and teachers must receive adequate training [43][44][45]. Instructors should be proficient in uploading multimedia, exchanging notes, assignments, and other learning resources on the LMS, and they should do so effectively. This includes the ability to switch between screens displaying different learning resources and write on an online whiteboard during a live online lecture session. They should also be able to send emails to students or other academic staff individually or to the entire class [44].

During online lectures, the instructor must maintain good posture and offer a camera view that does not demotivate learners. Even in online learning, facial expressions and eye contact play an essential role in providing an engaging camera view for the instructor [43]. Instructors may need to update their knowledge based on new technological trends to facilitate online learning effectively [46]. Participating in online courses can equip instructors with new skills and knowledge that can enhance their teaching, but their level of involvement may have positive or negative consequences [47].

Technological skills are crucial for teachers because they facilitate the transition of screen sharing between various digital instructional materials. In MOOCs, video lectures are the primary method of offering learning materials to students. Some video lecture material features the lecturer talking in front of a camera, with the whiteboard or blackboard displayed to students via their computer screens [42]. Students prefer short videos of around two to six minutes [48]. Using videos as a learning strategy is beneficial because they include a visual component, and viewers gain experiential value by seeing something they normally wouldn't see. However, it was observed that most learners who completed MOOC courses and earned certificates had accessed less than 80% of the video lecture materials [49].

The presence of a lecturer or instructor can trigger students to pay attention to the lecture, particularly gestures such as maintaining eye contact, which can foster learning [50]. In subjects with physical lab sessions, it is easier to assess student knowledge levels, guide them in practicals, and bridge any knowledge gaps. However, such behavior might be challenging in an online mode of instruction. Pakpour, Souto, and Schaffer (2021) have used interactive slides to teach Microbiology practicals to students in an online course. Instructors may use interactive teaching strategies, combined with reflective essays and discussion forums, to increase student engagement with the learning content [51]. Apps that facilitate the download of student attendance are preferable to manual attendance marking, which consumes necessary time that might otherwise be devoted to online live lectures.

A study monitoring the movement of students' mouse and eye movement during a course held online identified that the faster a student scrolls the mouse, the more familiar they are with the learning environment or the less engaged they are with the provided learning materials [52]. The more time students engage with the learning content, the more cognitive effort they exert [53]. Learners have diverse learning styles and come from various demographic backgrounds; therefore, teachers must use a wide range of teaching approaches to effectively reach the entire learner community [54].

During the transition from classroom learning to online learning, learners often encounter technical difficulties and may require training to use online learning platforms like Microsoft Teams and ZOOM [55]. A study found that high school students expected to use more technology and online learning for higher education while enrolled in tertiary education [56]. When both students and instructors are new to online education, it is essential to allocate time for them to become familiar with the technology. Learners must possess sufficient technical skills, including reading and downloading lecture materials, understanding the topics discussed and shared in discussion forums, delivering an online presentation, utilising social media and email, and using platforms that facilitate group activities. The key findings have been organized according to the CSE-SET framework, as illustrated in Fig. 5.

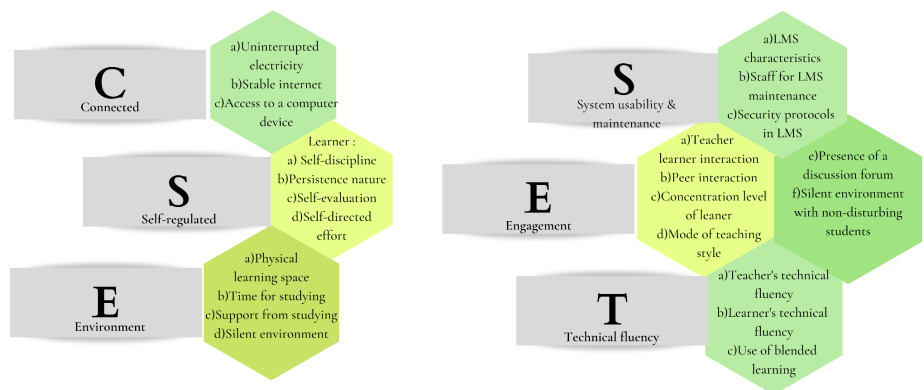

**Fig. 5.** The essentials for online learning : CSE-SET.

## 4 Conclusion

Online education provides a flexible mode of learning for learners and offers several benefits. However, it also poses challenges that require attention. This paper utilizes the work of other scholars on online learning to construct a framework that lists the essential and motivating factors needed for efficient and successful online learning. The findings of this study will be useful for online course designers to develop courses that will be highly appreciated by learners and teaching staff. Furthermore, academics can use the CSE-SET framework as shown in Fig. 5 to adopt better teaching styles suitable for online education.

**Declaration**

- The work received no financial support.

- The authors have no relevant financial or non-financial interests to disclose.
- Ethics approval is not applicable.
- Consent to participate is not applicable.
- Availability of data is not applicable.
- Code availability is not applicable.
- Authors' contributions : J.Dulangi Kanchana, is the primary author of this manuscript. She has read the recent literature and have constructed this manuscript under the supervision of the three supervisors; Dr.Gayashan Amarasinghe, Ms.Vishaka Nanayakkara and Dr.Amal Shehan Perera.